\documentclass[letterpaper, 10 pt, conference]{ieeeconf}  

\IEEEoverridecommandlockouts                              

\usepackage{cite}
\usepackage{amsmath,amssymb,amsfonts}
\usepackage{graphicx}
\usepackage{textcomp}
\usepackage{xcolor}

\usepackage{float}          
\usepackage{bm}             
\usepackage{orcidlink}      

\usepackage{subcaption}
\usepackage{booktabs}       
\usepackage{tabularx}       

\usepackage{algorithm}     
\usepackage{algpseudocode} 

\usepackage{tikz}                                
\usetikzlibrary{
    external,                     
    calc,                         
    pgfplots.groupplots,          
    arrows.meta,                  
    positioning,                  
    shapes.geometric,             
    shapes,                       
    plotmarks,                    
    fit,                          
    backgrounds,                  
    trees,                        
    decorations.pathreplacing,    
    tikzmark,                     
	arrows.meta,                  
	shapes.symbols,               
	patterns                      
}
\usepackage{pgfplots}
\pgfplotsset{compat=1.18}

\definecolor{mygreen}{RGB}{73,165,82}
\definecolor{myred}{RGB}{236,79,117}
\definecolor{myblue}{RGB}{10,30,123}
\definecolor{myviolet}{RGB}{138,43,227}
\definecolor{myorange}{rgb}{1.00000,0.64700,0.00000}%
\definecolor{myblack}{RGB}{0,0,0}       
\definecolor{mylightblue}{RGB}{0, 114, 189}
\definecolor{mydarkyellow}{RGB}{179,179,0}
\definecolor{mydarkorange}{RGB}{179,115,0}
\definecolor{mydarkblue}{RGB}{0,0,102}
\definecolor{mypurple}{RGB}{102,0,102}
\definecolor{mydarkred}{RGB}{153,0,0}
\definecolor{mydarkgreen}{RGB}{0,77,0}
\definecolor{mylightgray}{RGB}{179,179,179}

\tikzset{
  myCoordinateArrow/.style = {-{Stealth}},
  myCoordinateArrowBoth/.style={
        <->,
        >=stealth,
        shorten >=0pt,
        shorten <=0pt
    },
	mySignalArrow/.style = {-{Triangle}},
	mySignalArrowBoth/.style = {
    <->,                         
    >=Triangle,                  
    shorten >=0pt,               
    shorten <=0pt                
  }
}

\newcommand{\textopencircle}[1]{\textcolor{#1}{\raisebox{0.05ex}{\large$\circ$}}}
\newcommand{\textfilleddot}[1]{\textcolor{#1}{\raisebox{0.05ex}{\large$\bullet$}}}

\title{\LARGE \bf
Online and Offline Space-Filling Input Design for Nonlinear System Identification: A Receding Horizon Control-Based Approach
}

\author{Max Heinz Herkersdorf$^{1}$ and Oliver Nelles$^{1}$
\thanks{$^{1}$Max Heinz Herkersdorf 
and Oliver Nelles 
 are at Research Group Automatic Control - Mechatronics,
        University of Siegen, 57076, Siegen, Germany
        }}

\begin{document}

\maketitle
\thispagestyle{empty}
\pagestyle{empty}

\begin{abstract}
The effectiveness of data-driven techniques heavily depends on the input signal used to generate the estimation data. However, a significant research gap exists in the field of input design for nonlinear dynamic system identification. In particular, existing methods largely overlook the minimization of the generalization error, i.e., model inaccuracies in regions not covered by the estimation dataset.

This work addresses this gap by proposing an input design method that embeds a novel optimality criterion within a receding horizon control (RHC)-based optimization framework. The distance-based optimality criterion induces a space-filling design within a user-defined region of interest in a surrogate model's input space, requiring only minimal prior knowledge. Additionally, the method is applicable both online, where model parameters are continuously updated based on process observations, and offline, where a fixed model is employed.

The space-filling performance of the proposed strategy is evaluated on an artificial example and compared to state-of-the-art methods, demonstrating superior efficiency in exploring process operating regions.
\end{abstract}

\section{Introduction}
Contemporary applications of nonlinear system identification leverage advanced machine learning techniques to a great extent. The performance of these data-driven approaches is strongly dependent on the quality of the input signals used to generate estimation datasets \cite{nelles2020nonlinear, schoukens2019nonlinear, ljung1998system}. Consequently, alongside choosing appropriate model architectures and parameter estimation methods, careful design of input signals is essential. 

The primary objective of input design methodologies is to acquire precise and comprehensive information about the process behavior intended to be modeled. This is particularly challenging when capturing nonlinearities, which require data spanning the entire operational range \cite{deflorian2011design}, in combination with dynamic processes, whose operating points cannot be manipulated instantaneously and thus require both static and transient behaviors to be represented in the dataset \cite{herkersdorf2024optimized}. These difficulties are further exacerbated by real-world constraints, such as limited time, process constraints, and high measurement costs \cite{kosters2022optimization}.

\subsection{Localization in Related Work} \label{subsec:relatedWork}
The design of input signals, also referred to as excitation signals in system identification, can generally be categorized into model-based and model-free approaches.

\textit{Model-based} methods can be further subdivided into two primary categories: (i) Approaches that aim to minimize the variance of the parameter estimate by leveraging the Fisher information matrix (FIM) \cite{deflorian2010online, schrangl2020iterative, hametner2013optimal}. However, these methods are fundamentally limited by the assumption of an unbiased estimator, as required by the Cram\'er-Rao bound \cite{noam2009notes}. In practice, this assumption is often violated due to limited prior knowledge about the process. (ii) Approaches, typically following an active learning strategy, which aim to minimize the epistemic uncertainty of the underlying model \cite{buisson2020actively, sukhija2024optimistic, xie2024online}. Model-based input signal design can, in principle, generate highly informative datasets by directly optimizing process-focused optimality criteria. However, the success of these approaches heavily relies on the employed model's capability to accurately capture the process. Ensuring this capability in turn depends on prior knowledge -- information that is often unavailable.

\textit{Model-free} approaches represent a more traditional paradigm in input signal design. These methods require minimal prior knowledge, making them straightforward to implement and well-suited for identifying unknown processes. However, unlike model-based techniques, which can efficiently generate informative datasets when a suitable process model is available, model-free approaches lack a process-focused optimization and pursue heuristic strategies. Hence, they often require substantially larger datasets and, as a result, greater measurement costs to achieve the same level of information richness. State-of-the-art model-free excitation signal design methods include amplitude-modulated pseudo random binary signals (APRBS) \cite{nelles1995identification} and multisine \cite{pintelon2012system}.

The proposed method bridges the gap between model-free and model-based approaches by generating an excitation that optimizes the distribution of data points in a surrogate model’s input space. While a process model is still required for optimizing this distribution, the method can be considered "less model-based" than conventional model-based approaches. This distinction arises because the optimality criterion focuses on data distribution rather than intrinsic model properties, such as variance of the parameter estimate or uncertainty. As a result, systematic mismatches between the model architecture and the true process behavior do not introduce comparatively significant data bias, i.e., systematic distortions in the data distribution, such as undesired accumulations or gaps. Even a coarse surrogate of the process can yield effective results when optimizing the data distribution, as demonstrated in \cite{heinz2018excitation}.
Existing methods share a similar philosophy by employing data-distribution-centered optimality criteria. However, the proposed approach is unique in its flexibility: it can be executed offline, e.g., using only a coarse model that requires an approximate process time constant, as in \cite{herkersdorf2024optimized, heinz2018excitation, smits2021genetic}, or online in an adaptive modeling or active learning framework, where the employed model is continuously updated based on process observations, akin to \cite{vater2024differentiable}. Consequently, the method spans the spectrum from model-free to model-based, allowing users to leverage the respective advantages based on their specific requirements.
Other approaches that employ distribution-focused optimality criteria but rely a priori on a more precise process model -- and thus require extensive prior knowledge -- can be found in \cite{kiss2024space, liu2025space}. A localization of model-based and model-free excitation signal design methods for nonlinear system identification is illustrated in Fig. \ref{fig:ModelBased_vs_ModelFree}.
\begin{figure}[t]
	\centering
	{\begin{tikzpicture}[xscale=0.72, yscale=0.8]
	\def\t{0.056} 
	
	
	\draw[thick,mylightgray!60!black,fill=mylightgray!90,rounded corners=7,opacity=0.2]
   (-0.6,-2.3) rectangle (11.3,2.4);
   
   \draw[thick,mydarkgreen!60!black,fill=mydarkgreen!90,rounded corners=7,opacity=0.2]
	(2.6,-2.2) rectangle (8.1,-0.2);
    \node at ($(2.6,-2.2)!0.5!(8.1,-0.2) + (0,0.4)$) {novel contribution};
	\node[align=center] at ($(8.1,-2.2) + (-1,0.35)$){\footnotesize adaptive \\ \footnotesize modeling};
		\node[align=center] at ($(2.6,-2.2) + (1,0.35)$) {\footnotesize coarse \\ \footnotesize model};
		
\draw[thick,mydarkyellow!60!black,fill=mydarkyellow!90,rounded corners=7,opacity=0.2]
(0,0.2) rectangle (2.4, 2);
	\node[align=center] at ($(0,0.2)!0.5!(2.4, 2) $){\footnotesize APRBS \cite{nelles1995identification},  \\ \footnotesize multisine \cite{pintelon2012system}};
	
\draw[thick,mydarkorange!60!black,fill=mydarkorange!90,rounded corners=7,opacity=0.2]
(2.6,0.2) rectangle (5.2, 2);
\node[align=center] at ($(2.6,0.2)!0.5!(5.2, 2) $){\footnotesize Herkersdorf \cite{herkersdorf2024optimized}, \\ \footnotesize Heinz \cite{heinz2018excitation} , \\ \footnotesize Smits \cite{smits2021genetic}};

\draw[thick,mydarkblue!60!black,fill=mydarkblue!90,rounded corners=7,opacity=0.2]
(5.5,0.2) rectangle (8.1, 2);
\node[align=center] at ($(5.5,0.2)!0.5!(8.1, 2) $){\footnotesize Vater \cite{vater2024differentiable}, \\ \footnotesize Kiss \cite{kiss2024space}, \\ \footnotesize Liu \cite{liu2025space}};

\draw[thick,mypurple!60!black,fill=mypurple!90,rounded corners=7,opacity=0.2]
(8.3,0.2) rectangle (10.7, 2);
\node[align=center] at ($(8.3,0.2)!0.5!(10.7, 2) $){\footnotesize FIM-\\ \footnotesize focused \\ \footnotesize \cite{deflorian2010online, schrangl2020iterative, hametner2013optimal}};

\draw[thick,mydarkred!60!black,fill=mydarkred!90,rounded corners=7,opacity=0.2]
(8.3,-2.2) rectangle (10.7, -0.2);
\node[align=center] at ($(8.3,-2.2)!0.5!(10.7, -0.2) $){\footnotesize uncertainty-\\ \footnotesize focused \\ \footnotesize \cite{buisson2020actively, sukhija2024optimistic, xie2024online}};
	
\draw[myCoordinateArrowBoth, myblack, line width=2.5] 
(0,0) node[above, rotate=90] {model-free} 
-- (10.7,0) node[above, rotate=90, yshift=-15pt] {model-based};

\end{tikzpicture}}
	\caption{Localization of excitation signal design methods w.r.t. model-based and model-free approaches. }
	\label{fig:ModelBased_vs_ModelFree}
\end{figure}

\subsection{Space-Filling Design in Model Input Space}
The proposed method further distinguishes from existing approaches by introducing a novel optimality criterion embedded within a receding horizon control (RHC)-based optimization framework. In essence, the approach aims to generate a space-filling design within the relevant operating region of a surrogate model’s input space. This input space may correspond, for instance, to the joint input-state space, as in \cite{vater2024differentiable, kiss2024space, liu2025space}, or to the model regressor space, as considered in \cite{herkersdorf2024optimized, heinz2018excitation, smits2021genetic}. The fundamental premise is that the information an excitation signal provides about a process is inherently tied to the distribution it induces within the selected model space: if all relevant regions of a model's input space are sufficiently represented in the estimation dataset and the chosen model accurately predicts the process, comprehensive information is acquired about the operating region of interest. This consideration is particularly true for Markovian processes, where future states can be fully predicted given the knowledge of the current state \cite{van1992stochastic}. Consequently, the task of acquiring valuable information about the process behavior can be reframed as the challenge of sufficiently exciting the regions of interest within the model input space. 

The generalization error, i.e., model errors in regions of the operating space not covered by the estimation data, is one of the main sources of inaccuracies in nonlinear system identification \cite{nelles2020nonlinear, schoukens2019nonlinear}. A \textit{space-filling design} within the operating region of interest ensures comprehensive data collection of the process behavior intended to be modeled, minimizing knowledge gaps. Moreover, evenly distributed data mitigates data bias toward specific local operating regions, reducing the risk of overfitting \cite{mehrabi2021survey}. Consequently, space-filling excitation promotes the development of models that generalize well, contributing to robust and reliable predictions. 

\section{PROBLEM FORMULATION}
The primary objective of nonlinear system modeling is to minimize the discrepancy \begin{align}
e = y - \hat{y}
\label{eq:modelingGoal}
\end{align} between the actual process output $y$ and the predicted model output $\hat{y}$. Given a dynamic process, its dynamics can be represented by the discrete-time transfer function
\begin{align} 
	\hat{y}(k+1) = f\bigl(\underline{x}(k)\bigr) \hspace{3pt} \quad \forall \, k \in \mathbb{N}\ , 
	\label{eq:DiscreteTimeTransitionFunction} 
	\end{align} 
where $\underline{x} \in \mathbb{R}^{p}$ represents the $p$-dimensional model input and $k$ denotes the time step. The process dynamics depend on the constant sampling time $T_{s} \in \mathbb{R}_{+}$, i.e., the time interval between two discrete time steps. A common dynamics representation employed in nonlinear system identification is the Nonlinear AutoRegressive with eXogenous input (NARX) structure \cite{nelles2020nonlinear}. The structure makes use of $n_{u}$ past inputs and $n_{y}$ past outputs, leading to a model input or regression vector representation given by

\begin{align} 
	\begin{split}
		{\underline{x}}^{T}(k+1) =   &[ u_{1}(k), \ldots, u_{1}(k-m+1), \ldots, \\
		& u_{n_{u}}(k),  \ldots, u_{n_{u}}(k-m+1), \\
		& y_{1}(k), \ldots, y_{1}(k-m+1), \ldots, \\ 
		& y_{n_{y}}(k),  \ldots, y_{n_{y}}(k-m+1)] \quad \forall k \in \mathbb{N}  \ ,	\\ 								  		 
	\end{split}
	\label{eq:regressionVector}
\end{align}
where $m$ denotes the dynamic order.

In a data-driven modeling framework, $f(\cdot)$ in (\ref{eq:DiscreteTimeTransitionFunction}) is inferred from estimation data, making its accuracy contingent on the quality of the available dataset. Hence, the general objective of input design is to generate a dataset
\begin{align} 
	\begin{split}
		{\underline{U}}^{T} = [\underline{u}(1), \ldots, \underline{u}(N)]	\ , 		\hspace{9pt} N \in \mathbb{N}					  		 
	\end{split}
	\label{eq:excitationSignal}
\end{align}
of size $N$, where $\underline{u}^{T}(j) = [{u}_{1}(j), \ldots, {u}_{n_u}(j)]\  \forall j \in \{1, 2, \ldots, N\}$, which should capture as much information as possible about the process behavior intended to be modeled. This ensures that $f(\cdot)$ in (\ref{eq:DiscreteTimeTransitionFunction}) can be estimated accurately, thereby minimizing the error specified in (\ref{eq:modelingGoal}). In this contribution, the objective is pursued by optimizing the surrogate model's input space distribution according to a space-filling optimality criterion. For an NARX structure, this distribution corresponds to the regressor space distribution, which can be expressed for an excitation dataset as
\begin{align} \label{eq:regressorSpace}
	\begin{split}
		\underline{{X}}^{T}= [
		{\underline{x}}(1), {\underline{x}}(2), \ldots, {\underline{x}}(N) ] 	\ , 		\hspace{9pt} N \in \mathbb{N}\ \ . 
	\end{split}
\end{align}
The initial state of the process $\underline{x}(0)$, as well as the safe operating regions for the controllable inputs $\mathcal{U}$ and the regressor space $\mathcal{X}$, are assumed to be known, allowing the formulation of the following constraints to be satisfied: 
\begin{align}
	\begin{split} \label{eq:constraints}
		& \underline{{u}}(j) \in \mathcal{U}, \ \underline{{x}}(j) \in \mathcal{X} \hspace{3pt} \quad \forall \, j = \{1, 2, \ldots, N\} \ .
	\end{split}
\end{align}

\section{RHC-BASED SPACE-FILLING DESIGN} \label{sec:RHC-BASED SPACE-FILLING DESGIN}

In this section, the proposed input signal design methodology is introduced, and its key contributions are highlighted. First, we develop a receding horizon control (RHC)-based iterative optimization procedure, following a similar approach to \cite{gedlu2023online, hametner2013incremental}. Second, we formulate a novel optimality criterion that is embedded in the RHC-strategy and optimizes a surrogate model's input space distribution to achieve a space-filling design within the operating region of interest, thus ensuring comprehensive and balanced information acquisition about the process behavior intended to be modeled.
\subsection{Novel Optimality Criterion} \label{subsec:novel optimality criterion}
The proposed optimality criterion guides the information acquisition by directing the excitation toward the operating region of interest $\mathcal{C} \subseteq \mathcal{X}$, thereby exciting the process behavior to be modeled. This is achieved by uniformly distributing a supporting dataset $\underline{\Psi}$ within the corresponding relevant regions in the surrogate model's input space. Consequently, it follows that
\begin{align} \label{eq:supportingDataSet}
	\begin{split}
		 \underline{{\psi}}(j) \in \mathcal{C} \hspace{3pt}  \quad & \forall j = {1, 2, \ldots, N_{\Psi}}, \hspace{9pt} N_{\Psi} \in \mathbb{N}\ ,
	\end{split}
\end{align}
where $N_{\Psi}$ denotes the number of supporting points. The novel optimality criterion is defined as the sum of the nearest neighbor distances $d_{\mathrm{NN}}$ between each point in the supporting dataset and the data points of the excitation signal in the model input space:
\begin{align} \label{eq:OptimalityCriterion}
	\begin{split}
		 & J\bigl( \underline{\hat{X}}(k), \underline{{\Psi}}\bigr)  =  \sum_{j=1}^{N_{\Psi}}  d_{\mathrm{NN}}\bigl( {\underline{\psi}}(j), \underline{\hat{X}}(k) \bigr)  \\
		& \mathrm{with \, } \,
		 d_{\mathrm{NN}}\bigl( {\underline{\psi}}(j), \underline{\hat{X}}(k) \bigr)= \underset{1\leq i \leq k+L-1}{\min} ||\underline{\hat{x}}(i) - \underline{\psi}(j)||_{2}\, . 
	\end{split}
\end{align}
An NARX structure is selected as the model input space; therefore, the model input space distribution corresponds to the regressor space distribution $\underline{\hat{X}}(k)$, approximated by the surrogate model at time step $k$ of the signal. This approximation is necessary since access to future outputs -- required for executing the RHC-based optimization procedure, see Sec.\,\ref{subse:RHC-Like Optimization Procedure} -- as well as past outputs when running the proposed algorithm offline, is essential but unavailable.  Therefore, the yet unknown true process outputs $y$ are substituted with the surrogate model’s predictions $\hat{y}$. 

\subsection{RHC-Based Optimization Procedure} \label{subse:RHC-Like Optimization Procedure}The core principle of the RHC-based procedure lies in its iterative optimization. In each iteration, the optimal sequence $\tilde{\underline{U}}_{\mathrm{opt}}(k)$, starting at the current time step $k$ over a finite prediction horizon $L$, is determined as follows:
\begin{align} \label{eq:RHC}
	\begin{split}
		& \underline{ \tilde{U} }_{\mathrm{opt}}(k) =  \underset{ \underline{\tilde{U}} }{ \mathrm{arg\, min} } \ J \bigl(  \underline{\hat{X}}(k), \underline{ {\Psi}} \bigr) \\
		\mathrm{with} \
		& \underline{\hat{X}}(k) = \mathcal{M}_\theta([\underline{U}, \underline{\tilde{U}}]^{T}, \underline{\hat{x}}(0)) \\
		\mathrm{s. \, t.\ }  \
		& \underline{\tilde{u}}(j) \in \mathcal{U} \hspace{3pt} \quad \forall \, j = \{k, \ldots, k + L -1\} \\ 
		&\underline{\hat{x}}(j) \in \mathcal{\mathcal{X}}  \hspace{3pt} \quad \forall \, j = \{ k, \ldots,  k+L-1 \} \ .
	\end{split}
\end{align} 
Here, $\mathcal{M}_{\theta}$ represents the surrogate model, $\underline{U}$ denotes the previously optimized input data, and $\underline{\tilde{U}}$ refers to data points yet to be optimized. For a first-order dynamic system ($m=1$) with a single input ($n_{u}=1$) and a single output ($n_{y}=1$), the regressor space distribution constructed using the surrogate model at time step $k$ is defined by
\begin{align} \label{eq:surrogateRegressorSpaceAusgeschrieben}
	\underline{\hat{X}}(k) &=
	\begin{bmatrix}
		u(1) & \hat{y}(1) \\
		\vdots & \vdots \\
		u(k-1) & \hat{y}(k-1) \\
		\tilde{u}(k) & \hat{y}(k) \\
		\vdots & \vdots \\
		\tilde{u}(k+L-1) & \hat{y}(k+L-1) 
	\end{bmatrix}  \ ,
\end{align}
where $\hat{y}(j)\  \forall j \in \{1, 2, \ldots, k+L-1\}$ is predicted with the surrogate model $\mathcal{M}_{\theta}$.

Once $\underline{\tilde{U}}_{\mathrm{opt}}(k)$ is determined, its first data point $\tilde{\underline{u}}_{\mathrm{opt}}(k)$ is concatenated with the already optimized input data $\underline{U}$, and a forward time shift is performed, updating $k \rightarrow k+1$. This optimization procedure is iteratively repeated until the entire signal is generated, i.e., $k = N$. Pseudocode for the proposed procedure is presented in Algorithm 1.

\begin{algorithm}[h] \label{algo:RHCOptimization}
	\caption{The RHC-based optimization procedure}
	\textbf{Parameters:} Number of data points $N$, prediction horizon $L$, number of supporting points $N_{\Psi}$. \\
	\textbf{Initialization:} Safe operating region of the controllable inputs $\mathcal{U}$, safe operating region of the regressor space $\mathcal{X}$, supporting points $\underline{\Psi}$, surrogate model $\mathcal{M}_\theta$, initial state  $\underline{\hat{x}}(0)$. \\
	\begin{algorithmic}[H]
		\For{$k = 1, 2, \dots, N$} \Comment{Time steps.}
		\vspace{10pt}
		\State  $\hspace{0pt} \underline{ \tilde{U} }_{\mathrm{opt}}(k) =  \underset{ \underline{\tilde{U}} }{ \mathrm{arg\, min} } \ J \bigl(  \underline{\hat{X}}(k), \underline{ {\Psi}} \bigr)$
		\vspace{-8pt} \Comment{Get optimal \phantom{xxxxxxxxxxxxxxxxxxxxxxxxxxxxxxxxxxxx}sequence.}
		\vspace{-0pt}
		\begin{align*}
			\begin{split}
				\mathrm{with} \
				& \underline{\hat{X}}(k) = \mathcal{M}_\theta([\underline{{U}}, \underline{\tilde{U}}]^{T}, \underline{\hat{x}}(0)) \\
				\mathrm{s. \, t.\ }  \
				& \underline{\tilde{u}}(j) \in \mathcal{U} \hspace{3pt} \quad \forall \, j = \{k, \ldots, k + L-1 \} \\ 
				&\underline{\hat{x}}(j) \in \mathcal{\mathcal{X}}  \hspace{3pt} \quad \forall \, j = \{k, \ldots,  k+L-1 \} \\ 
			\end{split}
		\end{align*} 
		\State Apply $\underline{U} \leftarrow \underline{{U}} \cup  \underline{\tilde{u}}_{\mathrm{opt}}(k)$. 	\Comment{Append optimal \phantom{xxxxxxxxxxxxxxxxxxxxxxxxxxxxxxxx} data point.}
		\State Optimize $\mathcal{M}_{\theta}$ with \Comment{Only adaptive modeling.}					  
		\State data from $\underline{U}$.
		\State Go to $k \rightarrow k+1$. \Comment{Forward time-shift.}
		\EndFor
	\end{algorithmic}
\end{algorithm}
Figure~\ref{fig:RegressorSpaceWithSobol} provides an illustrative example of the optimization process at a given time step~$k$. The previously optimized data are indicated by \textcolor{mylightblue}{\rule[0.5ex]{0.7em}{0.5ex}}, while the optimal future sequence is highlighted in \textcolor{mygreen}{\rule[0.5ex]{0.7em}{0.5ex}}. Together, these form the regressor space distribution $\underline{{\hat{X}}}(k)$. Additionally, the uniformly distributed supporting dataset $\underline{\Psi}$, generated using Sobol sequences \cite{sobol1967distribution}, is depicted. To enhance visualization $N_{\Psi} < N$ is presented. In practice, however, it is advisable to choose the number of supporting points such that $N_{\Psi} \gg N$, allowing gaps to be efficiently targeted. By default, $N_{\Psi} = 5N$.
\begin{figure}[htbp]
    \centering

    \begin{subfigure}{0.22\textwidth}
        \centering
        \vspace{-4cm}
%
%
\definecolor{mycolor1}{rgb}{0.00000,0.44700,0.74100}%
\begin{tikzpicture}

\begin{axis}[%
width=0.7\textwidth,
height=0.3\textwidth,
at={(0\textwidth,0\textwidth)},
scale only axis,
xmin=1,
xmax=80,
xtick={ 1, 50, 80},
xticklabels={1, $k$, $k+L$},
xlabel style={font=\color{white!15!black}},
ymin=-0.02,
ymax=1.02,
ytick={  0, 0.5,   1},
ylabel style={font=\color{white!15!black}},
ylabel={$ u(k), \tilde{u}(k) $},
axis background/.style={fill=white}
]
\addplot [color=mylightblue, line width=1pt, forget plot]
  table[row sep=crcr]{%
1	0.411891287288054\\
2	0.80944913917873\\
3	0.0627244704994267\\
4	0.272707676899788\\
5	0.480531262064812\\
6	0.679746964251463\\
7	0.879973544885994\\
8	0.489436281485217\\
9	0.215380168926341\\
10	0.620605980535254\\
11	0.96031556602235\\
12	0.837903255867726\\
13	0.99999974312155\\
14	0.999999868972269\\
15	0.675422195361756\\
16	0.999999967078379\\
17	0.878364468403685\\
18	0.495667706140664\\
19	0.122486511626715\\
20	0.0735310217268102\\
21	0.0469950314461452\\
22	0.113835296315788\\
23	0.0507486229095991\\
24	0.164847393792392\\
25	0.919888862041992\\
26	0.256510236318404\\
27	0.415705458877177\\
28	0.962764366861907\\
29	0.730452626921664\\
30	0.927101154994022\\
31	0.609631305985568\\
32	0.725379034466893\\
33	0.499557446345293\\
34	0.878770810456702\\
35	0.945282563700309\\
36	0.933120067190402\\
37	0.843677335940851\\
38	0.815910795249261\\
39	0.706274484374401\\
40	0.945348307932254\\
41	0.814785188521522\\
42	0.22699241258201\\
43	0.827669104686273\\
44	0.90656494245711\\
45	0.605669856720335\\
46	0.932165411456447\\
47	0.370873679561902\\
48	0.586062225758283\\
49	0.244822075397459\\
50	0.303988033036755\\
};
\addplot [color=mygreen, line width=1pt, forget plot]
  table[row sep=crcr]{%
50	0.303988033036755\\
51	0.753164737717703\\
52	0.465783830228478\\
53	0.152714036210159\\
54	0.623522386640164\\
55	0.553272345287189\\
56	0.239227255132765\\
57	0.0324401326065116\\
58	0.208927159098402\\
59	0.115102674478165\\
60	0.24812827661508\\
61	0.139552205413704\\
62	0.0699713628913615\\
63	0.0770734384628636\\
64	0.11917611082213\\
65	0.154229894878362\\
66	0.229049751418381\\
67	0.0520598978613853\\
68	0.309669903264423\\
69	0.432145143761676\\
70	0.202522577514597\\
71	0.0418734995629527\\
72	0.603251328522919\\
73	0.185436723576199\\
74	0.013632639709578\\
75	0.918633428241361\\
76	0.0631454870352738\\
77	0.0836717477490948\\
78	0.383758027101802\\
79	0.787500175701866\\
80	0.67848774775502\\
};
\end{axis}
\end{tikzpicture}%
        \vspace{0.5em}
%
%
\definecolor{mycolor1}{rgb}{0.00000,0.44700,0.74100}%
\begin{tikzpicture}

\begin{axis}[%
width=0.7\textwidth,
height=0.3\textwidth,
at={(0\textwidth,0\textwidth)},
scale only axis,
xmin=1,
xmax=80,
xtick={ 1, 50, 80},
xticklabels={1, $k$, $k+L$},
xlabel style={font=\color{white!15!black}},
ymin=-0.02,
ymax=1.02,
ytick={  0, 0.5,   1},
ylabel style={font=\color{white!15!black}},
ylabel={$ \hat{y}(k)$},
axis background/.style={fill=white}
]
\addplot [color=mylightblue, line width=1pt, forget plot]
  table[row sep=crcr]{%
1	0.5\\
2	0.453689955576325\\
3	0.552473695458411\\
4	0.444502424749419\\
5	0.375048666819842\\
6	0.388385104211542\\
7	0.483353776649097\\
8	0.581189052517511\\
9	0.558592191179648\\
10	0.459608758873136\\
11	0.525576532160851\\
12	0.618939150959322\\
13	0.686904466683218\\
14	0.749523564228878\\
15	0.799618846732378\\
16	0.811476146155683\\
17	0.84918091575602\\
18	0.873755586785022\\
19	0.796391403951629\\
20	0.642752900212594\\
21	0.517229892671291\\
22	0.415613483491642\\
23	0.337625235760463\\
24	0.272091251314509\\
25	0.226121086014877\\
26	0.377550654071399\\
27	0.319336004212052\\
28	0.310718506602672\\
29	0.447153707449\\
30	0.538714374594378\\
31	0.627974064409771\\
32	0.65668461761043\\
33	0.705631136630215\\
34	0.66423787300332\\
35	0.725825256581081\\
36	0.778495642832558\\
37	0.820081949244913\\
38	0.848231884638961\\
39	0.868644534601656\\
40	0.872300705139332\\
41	0.895678899092659\\
42	0.906509919991315\\
43	0.73911593011743\\
44	0.782282223811202\\
45	0.821805428588884\\
46	0.810377221985466\\
47	0.845542831745985\\
48	0.715970897057\\
49	0.718255413728497\\
50	0.590482795049684\\
};
\addplot [color=mygreen, line width=1pt, forget plot]
  table[row sep=crcr]{%
50	0.590482795049684\\
51	0.496726886808197\\
52	0.581266459922897\\
53	0.544860734515321\\
54	0.443470204957353\\
55	0.513566498430068\\
56	0.541239627960005\\
57	0.448228222970325\\
58	0.359808790282883\\
59	0.299964200415032\\
60	0.245178554400467\\
61	0.212411447681525\\
62	0.176629504597943\\
63	0.144162544330824\\
64	0.118528016398367\\
65	0.100266377135288\\
66	0.0879000328472349\\
67	0.0844445114422135\\
68	0.0696036797567086\\
69	0.0810355778819158\\
70	0.127317377584273\\
71	0.113380320345398\\
72	0.0923174485236324\\
73	0.225927566380607\\
74	0.190793501955502\\
75	0.153131417962216\\
76	0.319097079874768\\
77	0.257820340554186\\
78	0.20977876570284\\
79	0.21132126201415\\
80	0.356600662221163\\
};
\end{axis}
\end{tikzpicture}%
        \caption{}
    \end{subfigure}
    \hspace{0.03\textwidth}
    \begin{subfigure}{0.22\textwidth}
        \centering
%
%
\definecolor{mycolor1}{rgb}{0.63500,0.07800,0.18400}%
\definecolor{mycolor2}{rgb}{0.00000,0.44700,0.74100}%
\definecolor{mycolor3}{rgb}{0.85000,0.32500,0.09800}%
\definecolor{mycolor4}{rgb}{0.92900,0.69400,0.12500}%
\begin{tikzpicture}

\begin{axis}[%
width=0.7\textwidth,
height=0.7\textwidth,
at={(0\textwidth,0\textwidth)},
scale only axis,
xmin=-0.02,
xmax=1.02,
xtick={  0, 0.5,   1},
xlabel style={font=\color{white!15!black}},
xlabel={$ u(k), \tilde{u}(k)$},
ymin=-0.02,
ymax=1.02,
ytick={  0, 0.5,   1},
ylabel style={font=\color{white!15!black}},
ylabel={$ \hat{y}(k)$},
axis background/.style={fill=white}
]
\addplot[only marks, mark=*, mark options={}, mark size=1.5000pt, color=mydarkred, fill=mycolor1, forget plot] table[row sep=crcr]{%
x	y\\
0	0\\
0.5	0.5\\
0.25	0.75\\
0.75	0.25\\
0.125	0.625\\
0.625	0.125\\
0.375	0.375\\
0.875	0.875\\
0.0625	0.9375\\
0.5625	0.4375\\
0.3125	0.1875\\
0.8125	0.6875\\
0.1875	0.3125\\
0.6875	0.8125\\
0.4375	0.5625\\
0.9375	0.0625\\
0.03125	0.53125\\
0.53125	0.03125\\
0.28125	0.28125\\
0.78125	0.78125\\
0.15625	0.15625\\
0.65625	0.65625\\
0.40625	0.90625\\
0.90625	0.40625\\
0.09375	0.46875\\
0.59375	0.96875\\
0.34375	0.71875\\
0.84375	0.21875\\
0.21875	0.84375\\
0.71875	0.34375\\
};
\addplot[only marks, mark=o, mark options={line width=1pt}, mark size=1.5000pt, draw=mylightblue, forget plot] table[row sep=crcr]{%
x	y\\
0.411891287288054	0.5\\
0.80944913917873	0.453689955576325\\
0.0627244704994267	0.552473695458411\\
0.272707676899788	0.444502424749419\\
0.480531262064812	0.375048666819842\\
0.679746964251463	0.388385104211542\\
0.879973544885994	0.483353776649097\\
0.489436281485217	0.581189052517511\\
0.215380168926341	0.558592191179648\\
0.620605980535254	0.459608758873136\\
0.96031556602235	0.525576532160851\\
0.837903255867726	0.618939150959322\\
0.99999974312155	0.686904466683218\\
0.999999868972269	0.749523564228878\\
0.675422195361756	0.799618846732378\\
0.999999967078379	0.811476146155683\\
0.878364468403685	0.84918091575602\\
0.495667706140664	0.873755586785022\\
0.122486511626715	0.796391403951629\\
0.0735310217268102	0.642752900212594\\
0.0469950314461452	0.517229892671291\\
0.113835296315788	0.415613483491642\\
0.0507486229095991	0.337625235760463\\
0.164847393792392	0.272091251314509\\
0.919888862041992	0.226121086014877\\
0.256510236318404	0.377550654071399\\
0.415705458877177	0.319336004212052\\
0.962764366861907	0.310718506602672\\
0.730452626921664	0.447153707449\\
0.927101154994022	0.538714374594378\\
0.609631305985568	0.627974064409771\\
0.725379034466893	0.65668461761043\\
0.499557446345293	0.705631136630215\\
0.878770810456702	0.66423787300332\\
0.945282563700309	0.725825256581081\\
0.933120067190402	0.778495642832558\\
0.843677335940851	0.820081949244913\\
0.815910795249261	0.848231884638961\\
0.706274484374401	0.868644534601656\\
0.945348307932254	0.872300705139332\\
0.814785188521522	0.895678899092659\\
0.22699241258201	0.906509919991315\\
0.827669104686273	0.73911593011743\\
0.90656494245711	0.782282223811202\\
0.605669856720335	0.821805428588884\\
0.932165411456447	0.810377221985466\\
0.370873679561902	0.845542831745985\\
0.586062225758283	0.715970897057\\
0.244822075397459	0.718255413728497\\
0.303988033036755	0.590482795049684\\
0.753164737717703	0.496726886808197\\
0.465783830228478	0.581266459922897\\
0.152714036210159	0.544860734515321\\
0.623522386640164	0.443470204957353\\
0.553272345287189	0.513566498430068\\
0.239227255132765	0.541239627960005\\
0.0324401326065116	0.448228222970325\\
0.208927159098402	0.359808790282883\\
0.115102674478165	0.299964200415032\\
0.24812827661508	0.245178554400467\\
};
\addplot[only marks, mark=o, mark options={line width=1pt}, mark size=1.5000pt, draw=mygreen, forget plot] table[row sep=crcr]{%
x	y\\
0.139552205413704	0.212411447681525\\
0.0699713628913615	0.176629504597943\\
0.0770734384628636	0.144162544330824\\
0.11917611082213	0.118528016398367\\
0.154229894878362	0.100266377135288\\
0.229049751418381	0.0879000328472349\\
0.0520598978613853	0.0844445114422135\\
0.309669903264423	0.0696036797567086\\
0.432145143761676	0.0810355778819158\\
0.202522577514597	0.127317377584273\\
0.0418734995629527	0.113380320345398\\
0.603251328522919	0.0923174485236324\\
0.185436723576199	0.225927566380607\\
0.013632639709578	0.190793501955502\\
0.918633428241361	0.153131417962216\\
0.0631454870352738	0.319097079874768\\
0.0836717477490948	0.257820340554186\\
0.383758027101802	0.20977876570284\\
0.787500175701866	0.21132126201415\\
0.67848774775502	0.356600662221163\\
};
\end{axis}
\end{tikzpicture}%
        \caption{}
    \end{subfigure}

    \caption{Illustrative example of the optimization process at time step~$k$: (a) input and corresponding output values, with previously optimized data in \textcolor{mylightblue}{\rule[0.5ex]{0.7em}{0.5ex}}, and the optimal future sequence in \textcolor{mygreen}{\rule[0.5ex]{0.7em}{0.5ex}}. (b) distribution of the regressor space $\underline{\hat{X}}(k)$ as approximated by the surrogate model in \textopencircle{mylightblue} / \textopencircle{mygreen}, together with the supporting dataset $\underline{{\Psi}}$ in \textfilleddot{mydarkred}.}
    \label{fig:RegressorSpaceWithSobol}
\end{figure}

\subsection{Implementation Details and Execution Modes}
The algorithm can be executed either \textit{online} within an adaptive modeling framework, where the surrogate model is continuously updated based on current process observations, or \textit{offline} with a fixed surrogate model. Possible scenarios for offline execution include the use of a coarse model that relies solely on an approximate process time constant, as in \cite{herkersdorf2024optimized, heinz2018excitation, smits2021genetic}, thus requiring only minimal process knowledge. Alternatively, if available, a white-box or gray-box model of the process can be employed as a fixed surrogate, similar to \cite{kiss2024space, liu2025space}.

Constraining the controllable inputs and the surrogate regressor space, as outlined in \,(\ref{eq:constraints}), is particularly beneficial for real-world applications, as it prevents operation in undesirable regions.

By default, a gradient-based optimizer is used to solve (\ref{eq:RHC}). However, more advanced optimization techniques can also be employed. To mitigate the risk of poor local optima, a multi-start approach is applied, initializing the optimization with both the solution from the previous iteration and randomly generated alternatives.

In this contribution, the optimization focuses on the surrogate regressor space distribution of the generated excitation signal. However, optimizing the joint input-state space obtained by a surrogate, as explored in \cite{vater2024differentiable, kiss2024space, liu2025space}, is also a viable approach. 

The following list summarizes the key advantages of the proposed excitation signal design method:
\begin{itemize}
	\item  Flexible execution in both online and offline modes, allowing for varying degrees of process adaptation while leveraging their respective advantages.
	\item Requires minimal prior knowledge for implementation.
	\item Incorporates constraints to prevent operation in undesirable regions.
	\item Piecewise $C^{1}$ differentiability, as well as $C^{0}$ differentiability, enabling efficient optimization via gradient-based techniques.
	\item No restriction to predefined signal shapes, such as sinusoidal or step-based signals.
\end{itemize}

\section{EVALUATION} \label{sec:EVALUATION}
In general, effective excitation signals lead to datasets that provide valuable information. A space-filling design within the operating region ensures comprehensive and balanced information acquisition about the process behavior intended to be modeled. Consequently, the primary focus of the evaluation section is to examine the space-filling performance of the proposed method, particularly in comparison to other state-of-the-art techniques.

Unfortunately, the concept of \textit{space-filling} design is not well-defined, particularly for dynamic processes. However, two key aspects are generally considered important: (i) ensuring that no region remains underrepresented, meaning that no gaps are left \cite{joseph2016space}, and (ii) achieving a distribution that covers the considered area as uniformly as possible \cite{lin2015latin}.
Hence, to evaluate the space-filling properties, two metrics are used. First, we examine the radius of the largest empty ball within the region of interest $\mathcal{C}$, defined as
\begin{align} \label{eq:RadiusLargestBall}
	\begin{split}
		& {R(}\underline{E}, \underline{{X}})  = \underset{1 \leq j \leq N_{E}}{\mathrm{max}}\bigl(  d_{\mathrm{NN}} (\underline{e}(j), \underline{{X}}) \bigr) \\
		& \mathrm{with} \ d_{\mathrm{NN}}\bigl( {\underline{e}}(j), \underline{{{X}}} \bigr)= \underset{1\leq i \leq N}{\min} ||\underline{{x}}(i) - \underline{e}(j)||_{2} \\
		&	\mathrm{s. \, t.\ }  \  \underline{e}(j) \in  \mathcal{C} \quad \forall \, j = \{ 1, 2, \ldots, N_{E} \} \ ,
	\end{split}
\end{align}
where $\underline{E}$ represents a uniformly distributed dataset, which in our investigations consists of $N_{E}=10000$ points. Second, to assess the uniformity of the space coverage, we employ the Jensen-Shannon divergence (JSD) with respect to the uniform distribution, given by
\begin{align} \label{eq:JSD}
	\mathrm{JSD}({X} \parallel {E}) &= \frac{1}{2} D_{\mathrm{KL}}({X} \parallel {M}) + \frac{1}{2} D_{\mathrm{KL}}({E} \parallel {M})\ ,
\end{align}
where $X$ denotes the data's probability distribution, $E$ the uniform distribution, ${M} = \frac{1}{2}({X} + {E})$ the mixture distribution, and $D_{\mathrm{KL}}$ the Kullback-Leibler divergence \cite{van2014renyi}.

Furthermore, both execution approaches presented--online execution with continuous updates of the model parameters based on current process observations, and offline execution with a fixed model, utilizing only an approximate process time constant as in \cite{herkersdorf2024optimized, heinz2018excitation, smits2021genetic}--are considered. 

A Hammerstein process is employed as example process in the evaluation, defined by:
 \begin{align} 	\label{eq:Hammerstein}
	\begin{split}	
		& y(k) = 0.2g(u(k-1))+0.8y(k-1), \\
		& g(x) =  \frac{\mathrm{atan}(8x-4)+\mathrm{atan}(4)}{2\mathrm{atan}(4)} \ .
	\end{split}
\end{align}
Since local model networks in a NARX structure trained with the local linear model tree (LOLIMOT) algorithm \cite{nelles2020nonlinear} can approximate the example process well, they are used as a surrogate model when following the online execution. In the offline approach, the surrogate model is a fixed first-order linear time-invariant (LTI) model with a time constant of $T = 5s$, which matches that of the example process, and a gain of $K = 1$. 

The prediction horizon is set to $L = 4T$, ensuring that the sequence currently optimized $\tilde{\underline{U}}(k)$ can access all operating regions of the process and thus effectively target gaps in the model input space distribution.

With the space-filling metrics, see \,(\ref{eq:RadiusLargestBall}) and \,(\ref{eq:JSD}), the process regressor space  
\begin{align} \label{eq:processRegSpace}
    \underline{X} = 
    \begin{bmatrix}
        u(1) & \cdots & u(N) \\
        y(1) & \cdots & y(N)
    \end{bmatrix}^{T}
\end{align} is examined. Here, $y(j) \  \forall j \in \{1, 2, \ldots, N\}$ are obtained from \,(\ref{eq:Hammerstein}). Although this space is generally not accessible in practice, in this artificial setting, it is very useful for evaluating the exploration of the process's operating regions.
 
\subsection{Theoretical Analysis} \label{subsec:theoreticalAnalysis}
In Fig.\,\ref{fig:OPTEXRegressorSpaceAndSignal}, a comparison is presented between the proposed method, which employs both the adaptive modeling strategy and the fixed LTI model as a surrogate. Even by visual inspection, it is evident that the process regressor space distribution obtained using the adaptive modeling strategy, $\underline{{X}}_{a}$, exhibits smaller gaps and, consequently, smaller underrepresented regions compared to its counterpart, $\underline{{X}}_{f}$, which was generated employing the fixed LTI. Hence, the proposed method using the adaptive approach can be regarded as superior in terms of space-filling, provided that the surrogate can adequately approximate the investigated process.
\begin{figure}[htbp]
    \centering

    \begin{subfigure}{0.22\textwidth}
        \centering
        \input{figures/Tikz/RegressorSpaceOPTEX_am}
        \caption{}
    \end{subfigure}
    \hspace{0.02\textwidth}
    \begin{subfigure}{0.22\textwidth}
        \centering
        \input{figures/Tikz/RegressorSpaceOPTEX_fm}
        \caption{}
    \end{subfigure}

    \vspace{0.5em}

    \begin{subfigure}{0.22\textwidth}
        \centering
%
%
%
\begin{tikzpicture}

\begin{axis}[%
width=0.7\textwidth,
height=0.35\textwidth,
at={(0\textwidth,0\textwidth)},
scale only axis,
xmin=1,
xmax=300,
xtick={  1, 300},
xlabel style={font=\color{white!15!black}},
xlabel={$ k $},
ymin=-0.02,
ymax=1.02,
ytick={  0, 0.5,   1},
ylabel style={font=\color{white!15!black}, yshift=-5pt},
ylabel={$ u_{a}(k) $},
axis background/.style={fill=white}
]
\addplot [color=mylightblue, line width=0.5pt, forget plot]
  table[row sep=crcr]{%
1	0.506731273119392\\
2	0.661018205388072\\
3	0.0905469656044213\\
4	0.345701593001158\\
5	0.0252105453836622\\
6	0.904789121016032\\
7	0.566739801686015\\
8	0.235175085221358\\
9	0.399330485139326\\
10	0.749477115920585\\
11	0.0920755903653076\\
12	0.480075578062857\\
13	0.921901680621822\\
14	0.634502690185953\\
15	0.162163268001594\\
16	0.542362781839473\\
17	0.930582077899303\\
18	0.757023647130125\\
19	0.57858293033606\\
20	0.92726974756813\\
21	0.692980697280184\\
22	0.940225014245625\\
23	0.802201862885301\\
24	0.999999936291865\\
25	0.944340597480826\\
26	0.448449055135447\\
27	0.579934730012845\\
28	0.999999916193018\\
29	0.999999929217459\\
30	0.665239246175775\\
31	0.999999984502583\\
32	0.999999418413843\\
33	0.802964755441445\\
34	0.910602505696433\\
35	0.10382263168141\\
36	0.882507916052482\\
37	0.891879605454967\\
38	0.266949563933339\\
39	0.130254666902731\\
40	0.263521370574024\\
41	0.807326788606848\\
42	0.681340564339375\\
43	0.435376212183158\\
44	0.873722681046532\\
45	0.774417816639518\\
46	0.310583052494839\\
47	0.0377153530804658\\
48	0.253823686964343\\
49	0.0197852385841579\\
50	0.222764759958045\\
51	0.0247245725602856\\
52	0.52249141798608\\
53	0.435121736390865\\
54	0.952742176283818\\
55	0.737193864527607\\
56	0.567706253564504\\
57	0.099335827102337\\
58	0.389422509462682\\
59	0.310037868233052\\
60	0.126038856662964\\
61	0.150390561070539\\
62	0.275227884803458\\
63	0.176182484801765\\
64	0.0541339147902443\\
65	0.265391958244923\\
66	0.00473008329233392\\
67	0.0436789534059404\\
68	0.414617625503422\\
69	0.791781020860304\\
70	0.328919099841772\\
71	0.829275136044094\\
72	0.841074470977906\\
73	0.0400715094968025\\
74	0.96914438989693\\
75	0.444984240486891\\
76	0.869825139028118\\
77	0.488714459732583\\
78	0.335682467102261\\
79	0.16534563826358\\
80	0.396288141671975\\
81	0.170461750011462\\
82	0.661839003683\\
83	0.888442968274957\\
84	0.766162318309749\\
85	0.965053284460722\\
86	0.856057358107718\\
87	0.498008891121984\\
88	0.629695142265081\\
89	0.757734616752685\\
90	0.911051373303319\\
91	0.395172727486385\\
92	0.218005462198893\\
93	0.0662664218812818\\
94	0.603561623120172\\
95	0.117832429941599\\
96	0.197060965173955\\
97	0.0434818819573465\\
98	0.0989757832204151\\
99	0.0353605303692969\\
100	0.207239911869166\\
101	0.0988986381532047\\
102	0.0263777101610357\\
103	0.151617643442546\\
104	0.0186699252414218\\
105	0.0880788504030638\\
106	0.203017348081462\\
107	0.312808658445126\\
108	0.557992047775648\\
109	0.437748761830171\\
110	0.661527934499844\\
111	0.315102160414546\\
112	0.849907403873111\\
113	0.691435146882061\\
114	0.702685352556298\\
115	0.956441439354679\\
116	0.971991546661087\\
117	0.999714634887589\\
118	0.738331889674498\\
119	0.940675863459234\\
120	0.050117174443686\\
121	0.949283174732547\\
122	0.827788133265427\\
123	0.972257811297344\\
124	0.999999975468068\\
125	0.999999923292929\\
126	0.570213571802791\\
127	0.846652153561887\\
128	0.999999995008886\\
129	0.973746546518169\\
130	0.937818381757746\\
131	0.703784367400798\\
132	0.217595228268131\\
133	0.64196218502423\\
134	0.854711515590116\\
135	0.954907190679783\\
136	0.969125304207135\\
137	0.902269662342569\\
138	0.958968628428119\\
139	0.34794675157921\\
140	0.926617709780446\\
141	0.491526985298619\\
142	0.750148388412135\\
143	0.151404769815554\\
144	0.879930193427066\\
145	0.440355569806253\\
146	0.542777520934904\\
147	0.0491918806486263\\
148	0.982813639341919\\
149	0.823753621225586\\
150	0.820450169775989\\
151	0.787721539446308\\
152	0.0441596489102742\\
153	0.157775928382053\\
154	0.223476992298377\\
155	0.577592350619617\\
156	0.338673440321794\\
157	0.772338532779439\\
158	0.371913115700873\\
159	0.941814616498781\\
160	0.351590767808229\\
161	0.4298092513684\\
162	0.152191710356334\\
163	0.250458062024788\\
164	0.260379181826887\\
165	0.212621817992638\\
166	0.0553015241652549\\
167	0.118203552765888\\
168	0.14377100535276\\
169	0.189846786152949\\
170	0.052720032324439\\
171	0.0885542575303513\\
172	0.924768424555359\\
173	0.556058973120088\\
174	0.342898541382555\\
175	0.0635026256635746\\
176	0.239541610422435\\
177	0.956452105065363\\
178	0.643526396791033\\
179	0.489679451569672\\
180	0.0890614274410357\\
181	0.442206446313303\\
182	0.803721924073966\\
183	0.658828880520484\\
184	0.907368435814165\\
185	0.690178771229683\\
186	0.900486585216092\\
187	0.3787301429954\\
188	0.622045904287422\\
189	0.36555083485776\\
190	0.827831663772988\\
191	0.502357719116716\\
192	0.239463753585969\\
193	0.813843985703944\\
194	0.889816076196618\\
195	0.907702782272044\\
196	0.217198779227893\\
197	0.913316958605245\\
198	0.932606730926578\\
199	0.753428704901421\\
200	0.458437418029586\\
201	0.821568313857708\\
202	0.0843664289971738\\
203	0.587880917121616\\
204	0.4039662379741\\
205	0.482025524296079\\
206	0.212118957079748\\
207	0.189873410532905\\
208	0.204124712492126\\
209	0.179538876805208\\
210	0.740912047827781\\
211	0.972449837113884\\
212	0.296013119777229\\
213	0.246411588415215\\
214	0.693404889403531\\
215	0.769436178412423\\
216	0.0936210690999548\\
217	0.11936873722395\\
218	0.0729200391763813\\
219	0.474949437927601\\
220	0.603019386218496\\
221	0.0764873694601824\\
222	0.10857505831241\\
223	0.126468502408751\\
224	0.219864159737554\\
225	0.187011001009617\\
226	0.0716512362664192\\
227	0.0354512860237468\\
228	0.111136966315889\\
229	0.0335370223191699\\
230	0.238080993033793\\
231	0.129271426094775\\
232	0.760662171004868\\
233	0.136483863963423\\
234	0.314202727802282\\
235	0.724914699829335\\
236	0.809569343145249\\
237	0.868156515581534\\
238	0.610720342749075\\
239	0.935036487082551\\
240	0.473008999072968\\
241	0.8026285882097\\
242	0.777541086118326\\
243	0.289978953724933\\
244	0.95344644677514\\
245	0.730717980475097\\
246	0.981278839951582\\
247	0.849335832091343\\
248	0.885824687355957\\
249	0.77039822589178\\
250	0.645724632903062\\
251	0.328869612936633\\
252	0.15452036733508\\
253	0.734530096579846\\
254	0.66356272842418\\
255	0.895917427565604\\
256	0.543896840917853\\
257	0.59789516344581\\
258	0.0339045921248769\\
259	0.654296571896752\\
260	0.746889204893113\\
261	0.688897653521708\\
262	0.961430778525967\\
263	0.712474052049631\\
264	0.861196113582981\\
265	0.915378101774387\\
266	0.926795089460898\\
267	0.924242362715137\\
268	0.850511275822194\\
269	0.988308752736297\\
270	0.868259050366257\\
271	0.940940207962564\\
272	0.636644171486782\\
273	0.76710966607387\\
274	0.515354271656504\\
275	0.15651195736895\\
276	0.187076964761223\\
277	0.509893909077703\\
278	0.945424417034441\\
279	0.299666147704266\\
280	0.843898050916446\\
281	0.693128797012482\\
282	0.544047930358949\\
283	0.0319064943073937\\
284	0.651603924828572\\
285	0.0864542680003779\\
286	0.153861037429021\\
287	0.487855534341524\\
288	0.34468355803253\\
289	0.966432451869916\\
290	0.958618618019887\\
291	0.329239464681118\\
292	0.0693066730893268\\
293	0.026006080118428\\
294	0.361836155213519\\
295	0.113173323993844\\
296	0.316790863333471\\
297	0.199677432694672\\
298	0.390300671807264\\
299	0.145828355379842\\
300	0.591857730819187\\
};
\end{axis}
\end{tikzpicture}%
        \caption{}
    \end{subfigure}
    \hspace{0.02\textwidth}
    \begin{subfigure}{0.22\textwidth}
        \centering
%
%
%
\begin{tikzpicture}

\begin{axis}[%
width=0.7\textwidth,
height=0.35\textwidth,
at={(0\textwidth,0\textwidth)},
scale only axis,
xmin=1,
xmax=300,
xtick={  1, 300},
xlabel style={font=\color{white!15!black}},
xlabel={$ k $},
ymin=-0.02,
ymax=1.02,
ytick={  0, 0.5,   1},
ylabel style={font=\color{white!15!black}, yshift=-5pt},
ylabel={$ u_{f}(k) $},
axis background/.style={fill=white}
]
\addplot [color=mylightblue, line width=0.5pt, forget plot]
  table[row sep=crcr]{%
1	0.136381369936296\\
2	0.160038624606691\\
3	0.413031485418531\\
4	0.67513563593945\\
5	0.109628717388081\\
6	0.880599252719808\\
7	0.582095058245358\\
8	0.330863307822098\\
9	0.893123862891285\\
10	0.646595995498909\\
11	0.928262602547409\\
12	0.961188195082616\\
13	0.7637612491333\\
14	0.905397502361677\\
15	0.999999941807847\\
16	0.99999987316491\\
17	0.999999984617\\
18	0.515031970754418\\
19	0.999999996165616\\
20	0.916766785505819\\
21	0.999999958963886\\
22	0.755960022553626\\
23	0.302425134535793\\
24	0.0802788452647989\\
25	0.574482568766191\\
26	0.851380668097435\\
27	0.979067550122914\\
28	0.702920500944942\\
29	0.831727836090185\\
30	0.441875048764795\\
31	0.358147938744442\\
32	0.678719181473609\\
33	0.229623917514077\\
34	0.0567008981582433\\
35	0.224567870074241\\
36	0.309787886446179\\
37	0.471171732824845\\
38	0.0321494718422794\\
39	0.212666805589042\\
40	0.0197340537766616\\
41	6.7397587011732e-08\\
42	0.27147838352309\\
43	0.167744603737788\\
44	0.0307380602713375\\
45	0.428421070991707\\
46	0.2155031841253\\
47	0.580622450176244\\
48	0.113445054035716\\
49	0.35024462062746\\
50	0.459305032709157\\
51	0.740809721560852\\
52	0.964897651774982\\
53	0.563319273775638\\
54	0.866889333337754\\
55	0.922235021242822\\
56	0.999999905970001\\
57	0.958412676555467\\
58	0.999999943284173\\
59	0.999999960195944\\
60	0.999999990979244\\
61	0.998273337653421\\
62	0.649698598031782\\
63	0.999999998456188\\
64	0.870750732599313\\
65	0.960458145507832\\
66	0.114431524523822\\
67	0.615140236048099\\
68	0.21530875136294\\
69	0.310270620887459\\
70	0.602612854705428\\
71	0.793244059728589\\
72	0.418351601988539\\
73	0.502393698324872\\
74	0.710927448651026\\
75	0.908394320464959\\
76	0.0783209855123342\\
77	0.230448749693087\\
78	0.0746660772379541\\
79	0.842920154107628\\
80	0.021164573537069\\
81	0.235251716433161\\
82	0.131610987005936\\
83	0.402158221867226\\
84	0.160091257873025\\
85	0.0582360261833664\\
86	0.0342515007682236\\
87	0.773949749359399\\
88	0.84664354215734\\
89	0.933358946477309\\
90	0.369502838771574\\
91	0.318165413375603\\
92	0.270518868956845\\
93	0.17679714163633\\
94	0.184358534677294\\
95	0.284992676429697\\
96	0.506740630332238\\
97	0.660106071640452\\
98	0.523636683245347\\
99	0.06788830094907\\
100	0.0431024697055842\\
101	1.23586102668698e-06\\
102	6.15198276520256e-09\\
103	3.89727628305521e-07\\
104	0.0703998929418616\\
105	8.87465058190453e-08\\
106	4.44218663524441e-09\\
107	0.0661992391990619\\
108	0.233445576655063\\
109	0.798747519098944\\
110	0.056243798004275\\
111	1.2955294028646e-07\\
112	0.00458905882782056\\
113	0.320487886029631\\
114	0.520566767887171\\
115	8.01236088284317e-08\\
116	1.14632535462644e-08\\
117	0.946967294373444\\
118	0.950636782238086\\
119	0.589907524061801\\
120	0.734653116053446\\
121	0.750502371680269\\
122	0.300655987657499\\
123	0.45396380321244\\
124	0.806023869166497\\
125	0.144016287797933\\
126	0.9730221742698\\
127	0.558800673582681\\
128	0.682763067543443\\
129	0.977341298016539\\
130	0.0273055099262023\\
131	0.863650922825517\\
132	0.821233991398176\\
133	0.63520369360646\\
134	0.153199552762269\\
135	0.894884983072518\\
136	0.880502791944258\\
137	0.744016920269704\\
138	0.861826769776245\\
139	0.404037030067442\\
140	0.967742597596189\\
141	0.909739789532224\\
142	0.542886370455183\\
143	0.999999978163675\\
144	0.964859142237269\\
145	0.999246747743141\\
146	0.41378159615417\\
147	0.888993322592084\\
148	0.83146865731871\\
149	0.647112390386333\\
150	0.792204587461921\\
151	0.196334350011567\\
152	0.988992241343262\\
153	0.778974544938684\\
154	0.878655625625532\\
155	0.307759542613047\\
156	0.81411352840669\\
157	0.145279551749333\\
158	0.759804439529256\\
159	0.497891083359869\\
160	0.176939328174961\\
161	0.63354745156096\\
162	0.1904397076035\\
163	0.517614769479455\\
164	0.294486583521039\\
165	0.430390596678444\\
166	0.103590581763879\\
167	0.789287670150306\\
168	0.397420760154944\\
169	0.800799409157407\\
170	0.0883039400519045\\
171	0.881166925503735\\
172	0.0262226436357011\\
173	0.365525897682127\\
174	0.1157750795954\\
175	0.704189559251961\\
176	0.267204003471681\\
177	0.086991139049453\\
178	0.906906689882419\\
179	0.191045226692644\\
180	0.725607257994883\\
181	0.150013268279803\\
182	0.148925776333686\\
183	0.0533415249948292\\
184	0.142850660040447\\
185	0.20247760148391\\
186	0.258995748408124\\
187	2.58661895145273e-07\\
188	0.664857203808484\\
189	0.330286712064062\\
190	0.377969223254688\\
191	0.669108306347793\\
192	0.0613342925304936\\
193	0.10362009192353\\
194	0.257867862189387\\
195	0.873815677565139\\
196	0.910890501847726\\
197	0.469744786205063\\
198	0.864127822153538\\
199	0.51140632982421\\
200	0.411106416222757\\
201	0.781543500323514\\
202	0.868833226911778\\
203	0.370083834689267\\
204	0.124766113520661\\
205	0.248275202866745\\
206	0.297011430778007\\
207	0.664836186981894\\
208	0.150801940020298\\
209	0.280399381629361\\
210	0.41286430878384\\
211	0.252333369118476\\
212	0.55216729413714\\
213	0.768525220876977\\
214	0.847008932430113\\
215	0.902463916103951\\
216	0.714331801493545\\
217	0.630915952907562\\
218	0.21077939854987\\
219	0.970808762866873\\
220	0.190140252399488\\
221	0.599461738621566\\
222	0.811924695222624\\
223	0.588790875871898\\
224	0.926981322865741\\
225	0.927521538262098\\
226	0.759938039243304\\
227	0.965975672441682\\
228	0.732594770077593\\
229	0.931631746428891\\
230	0.0360312373467132\\
231	0.838835590594376\\
232	0.307571435556852\\
233	0.256226655304671\\
234	0.860650178003679\\
235	0.954617649256603\\
236	0.830603970475257\\
237	0.0483177902482006\\
238	0.459374955967046\\
239	0.941210831841185\\
240	0.948008558719639\\
241	0.683589203956294\\
242	0.998790794808105\\
243	0.999470003443669\\
244	0.999999669541496\\
245	0.999991595157198\\
246	0.999999965814535\\
247	0.482581025718778\\
248	0.579618358018019\\
249	0.598484115222392\\
250	0.497293066852814\\
251	0.630268244103277\\
252	0.555838933905475\\
253	0.598926088150039\\
254	0.777450069979139\\
255	0.784103330576934\\
256	0.938369124291516\\
257	0.675047158911196\\
258	0.94599341373967\\
259	0.922633771044434\\
260	0.822675757530119\\
261	0.239171861366071\\
262	0.269897861812846\\
263	0.749297146731406\\
264	0.737451508360792\\
265	0.718789832169718\\
266	0.203318443572166\\
267	0.838763399627544\\
268	0.801976240786378\\
269	0.465039285524333\\
270	0.0823568336457362\\
271	0.827294569126525\\
272	0.352936540576115\\
273	0.846744630137385\\
274	0.0270419931322547\\
275	0.182398522329212\\
276	0.569765992329793\\
277	0.0283708836624478\\
278	0.105079663304922\\
279	0.137820842892094\\
280	0.488667470724759\\
281	0.0278338497718446\\
282	0.0324103138307968\\
283	0.121470811030676\\
284	8.88771973478263e-07\\
285	0.183512376694176\\
286	1.4234855601331e-08\\
287	0.125029053453197\\
288	1.60144661917527e-07\\
289	0.562180358659658\\
290	0.375747101766241\\
291	0.343656435361869\\
292	0.180048734382146\\
293	0.512523513146766\\
294	0.544285337079013\\
295	0.0848785004943438\\
296	0.0909081584504578\\
297	0.128226718740522\\
298	0.305038650528145\\
299	0.119086316964933\\
300	0.441253537611855\\
};
\end{axis}
\end{tikzpicture}%
        \caption{}
    \end{subfigure}

    \caption{Illustrations of the process regressor space distribution $\underline{{X}}_{a}$, see (\ref{eq:processRegSpace}), and the corresponding signal trajectory $\underline{u}_{a}$, generated using the proposed method with adaptive modeling, are shown in (a) and (c). Their counterparts, obtained using the fixed LTI model, $\underline{{X}}_{f}$ and $\underline{u}_{f}$, are shown in (b) and (d).}
    \label{fig:OPTEXRegressorSpaceAndSignal}
\end{figure}

\subsection{Space-Filling Performance} \label{subsec:Space-Filling Performance}
In this section, the space-filling performance of the proposed method, using both the adaptive strategy and a fixed first-order LTI model as a surrogate, is compared to state-of-the-art techniques. First, a comparison is made to the incremental dynamic space-filling design (IDS-FID) approach presented in \cite{herkersdorf2024optimized}. In the evaluation, the IDS-FID algorithm employs the same LTI model as the proposed method. Second, APRBS with minimum holding time $T_{H}=1s$ are used for comparison. For each excitation signal design method, 50 signals are generated, each containing a dataset of size $N = 300$. An overview of the investigated signals is provided in Table \ref{tab:excitationMethods}. The sampling time is set to $T_{s}=1s$. 
\begin{table}[h]
	\centering
	\begin{tabularx}{0.45\textwidth}{l X} 
		\toprule
		Excitation method & Properties \\
		\midrule
		$\underline{U}_{a}$ & Proposed method employing adaptive modeling. \\
		$\underline{U}_{f}$ & Proposed method employing the fixed LTI model with $T=5s, K = 1$. \\
		$\underline{U}_{\mathrm{IDSFID}}$ & Excitation generated using the IDS-FID algorithm with hyperparameter values $\lambda \in (0, 0.5)$, resulting in a spectrum ranging from predominantly static to rather transient excitation. Employs the fixed LTI, as well. \\
		$\underline{U}_{\mathrm{APRBS}}$ & APRBS with $T_{H}=1s$.\\
		\bottomrule
	\end{tabularx}
	\caption{Properties of the investigated excitation signal methods. For each method, 50 excitation signals of size $N=300$ were generated.}
	\label{tab:excitationMethods}
\end{table}

\begin{figure}[h]
	\centering
	\includegraphics[width=0.45\textwidth, trim=0 0 0 0, clip]{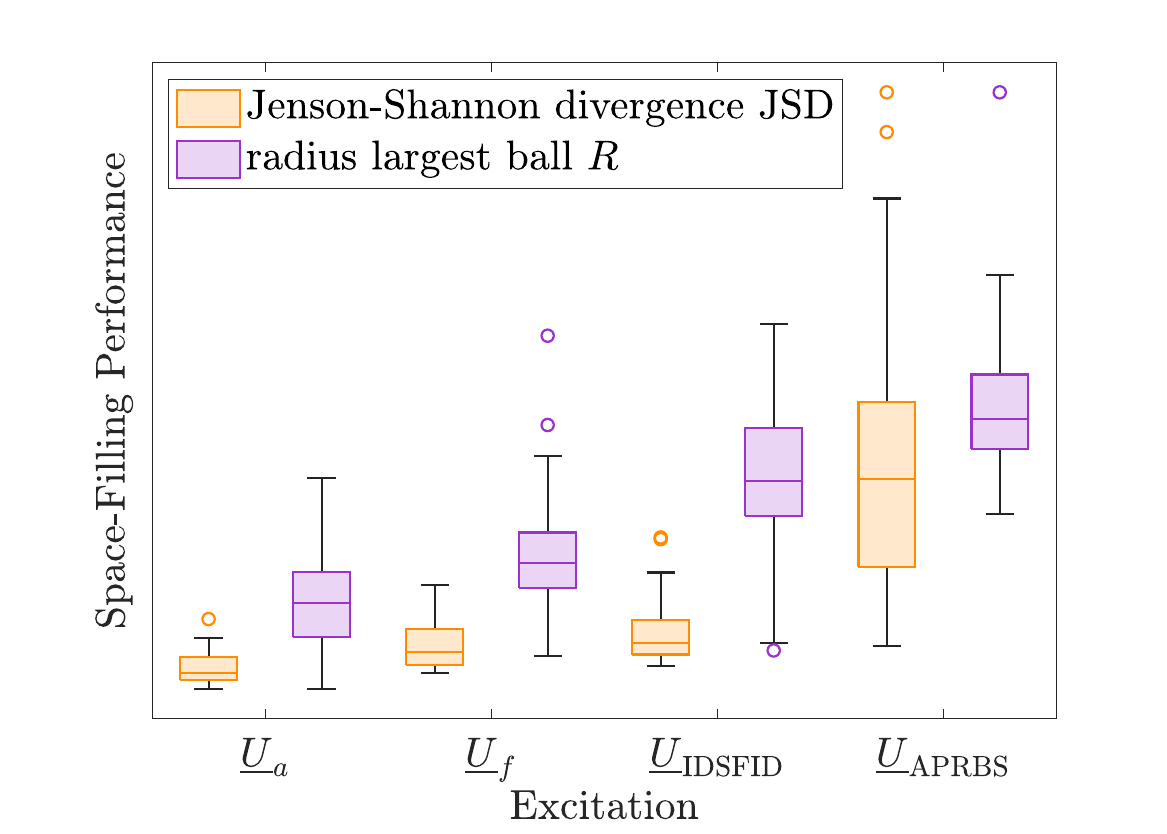}
	\caption{Boxplot of the space-filling performance of the regressor space distribution resulting from different excitation signal design methods. For each method, 50 excitation signals of size $N=300$ were generated.}
	\label{fig:space_filling}
\end{figure}

Figure \ref{fig:space_filling} presents the results of the investigations. The proposed method clearly outperforms state-of-the-art approaches in both adaptive modeling execution and employing the fixed surrogate model. Since the IDS-FID strategy utilizes the same surrogate as the proposed method in offline mode, it can be concluded that the optimization procedure of the proposed method is superior in terms of space-filling.

Figure~\ref{fig:LargestBallProgressSequence} shows how the radius of the largest ball ${R}(k)$ evolves as the dataset size increases. Displayed are the median-performing signals of the investigations shown in Fig.\,\ref{fig:space_filling}. Initially, the performance of $\underline{u}_{a}$ and $\underline{u}_{f}$ is similar. However, as the dataset size increases, the proposed method employing the adaptive modeling approach demonstrates superiority. This improvement is attributed to the continuous updating of model parameters, which refines the model as more process observations are collected, thereby enabling efficient targeting of gaps in the data distribution. The same investigations as in Fig.\,\ref{fig:LargestBallProgressSequence} were conducted using the JSD. However, since the results are similar, they are not explicitly presented here.
\begin{figure}[h]
	\centering
	{\input{figures/Tikz/LBProgressSequence}}
	\caption{Progress of the radius of the largest ball ${R}$ as a function of the dataset size for different excitation signal design methods. The median-performing signals of the investigations shown in Fig.\,\ref{fig:space_filling} are displayed.}
	\label{fig:LargestBallProgressSequence}
\end{figure}

\section{Discussion and Outlook}
The proposed method addresses the generalization error by inducing a space-filling distribution in a surrogate model's input space, achieving superior performance over existing input signal design approaches. Future work will assess the method on more complex and realistic processes to further validate these results.

The results further highlight that the effectiveness of model-based input design is closely linked to the surrogate model's quality. Although the approach is less dependent on intrinsic model properties, such as parameter estimate variance and uncertainty, than traditional methods, a systematic analysis of the interplay between surrogate model accuracy and the resulting input signal quality is required and will be pursued.

Further research will focus on:
\begin{itemize}
    \item Systematic comparison of the employed spaces, specifically the joint input-state space and the regressor space.
    \item Assessing whether distributions other than space-filling designs may be beneficial under specific conditions, such as the presence of solely local nonlinearities.
    \item Analyzing computational complexity, especially in the context of online execution, the number of supporting points, and the length of the prediction horizon.
    \item Evaluating the impact of the number of supporting points and the length of the prediction horizon on the input signal quality.
\end{itemize}

\bibliographystyle{IEEEtran}
\bibliography{IEEEabrv,myRefsMaxHerk}

\end{document}